\documentclass[conference]{IEEEtran}
\usepackage{graphicx}
\usepackage{caption}
\usepackage{subcaption}
\usepackage{float}

\ifCLASSINFOpdf
\else
\fi
\begin{document}
%
\title{Design and Evaluation of LAr Trigger Digitizer Board in ATLAS Phase-I Upgrade}



%
\author{\IEEEauthorblockN{Kai Chen\IEEEauthorrefmark{1},
Hucheng Chen\IEEEauthorrefmark{1},
Mauro Citterio\IEEEauthorrefmark{3},
Herve Deschamps\IEEEauthorrefmark{2}, 
Aude Grabas\IEEEauthorrefmark{2},
Stefano Latorre\IEEEauthorrefmark{3},\\
Massimo Lazzaroni\IEEEauthorrefmark{3},
Hongbin Liu\IEEEauthorrefmark{1},
Phillippe Schwemling\IEEEauthorrefmark{2},
Stefan Simion\IEEEauthorrefmark{4}},
Hao Xu\IEEEauthorrefmark{1} and
Heling Zhu\IEEEauthorrefmark{5},

\IEEEauthorblockA{\IEEEauthorrefmark{1}Physics Department,
Brookhaven National Laboratory, Upton, NY, USA}
\IEEEauthorblockA{\IEEEauthorrefmark{2}CEA-Saclay, Paris, France}
\IEEEauthorblockA{\IEEEauthorrefmark{3}INFN Milano, Milan, Italy}
\IEEEauthorblockA{\IEEEauthorrefmark{4}LAL, Orsay, France}
\IEEEauthorblockA{\IEEEauthorrefmark{5}University of Science and Technology of China, Hefei, China}
}


\maketitle

\begin{abstract}
The LHC upgrade is planned to enhance the instantaneous luminosity during Run 3 from 2021 through 2023. The Phase-I upgrade of the trigger readout electronics for the ATLAS Liquid Argon (LAr) Calorimeters will be installed during the second long shutdown of LHC in 2019-2020. In this upgrade, the so-called super cells are introduced to provide higher granularity, higher resolution and longitudinal shower shape information from the LAr calorimeters to the level-1 trigger processors. A new LAr Trigger Digitizer Board (LTDB) will manipulate and digitize 320 channels of super cell signals, and transmit it via 40 fiber optical links to the back end where data are further processed and transmitted to the trigger processors. Five pairs of bidirectional GBT links are used for slow control from the Front-end LInks eXchange (FELIX) in the ATLAS TDAQ system. LTDB also outputs 64 summed analog signals to the current Tower Builder Board via the new baseplane. A test system is developed to test all functions of the LTDB and carry out the performance measurement. A back end PCIe card is designed which has the circuit to interface to the ATLAS trigger, time and control system. It can control the generation of injection signals to the LTDB for performance test. It also configures and calibrate all ASICs on the LTDB.
\end{abstract}



%
\IEEEpeerreviewmaketitle

\section{Introduction}

\begin{figure}[h!]
	\centering
	\includegraphics[width=0.95\linewidth]{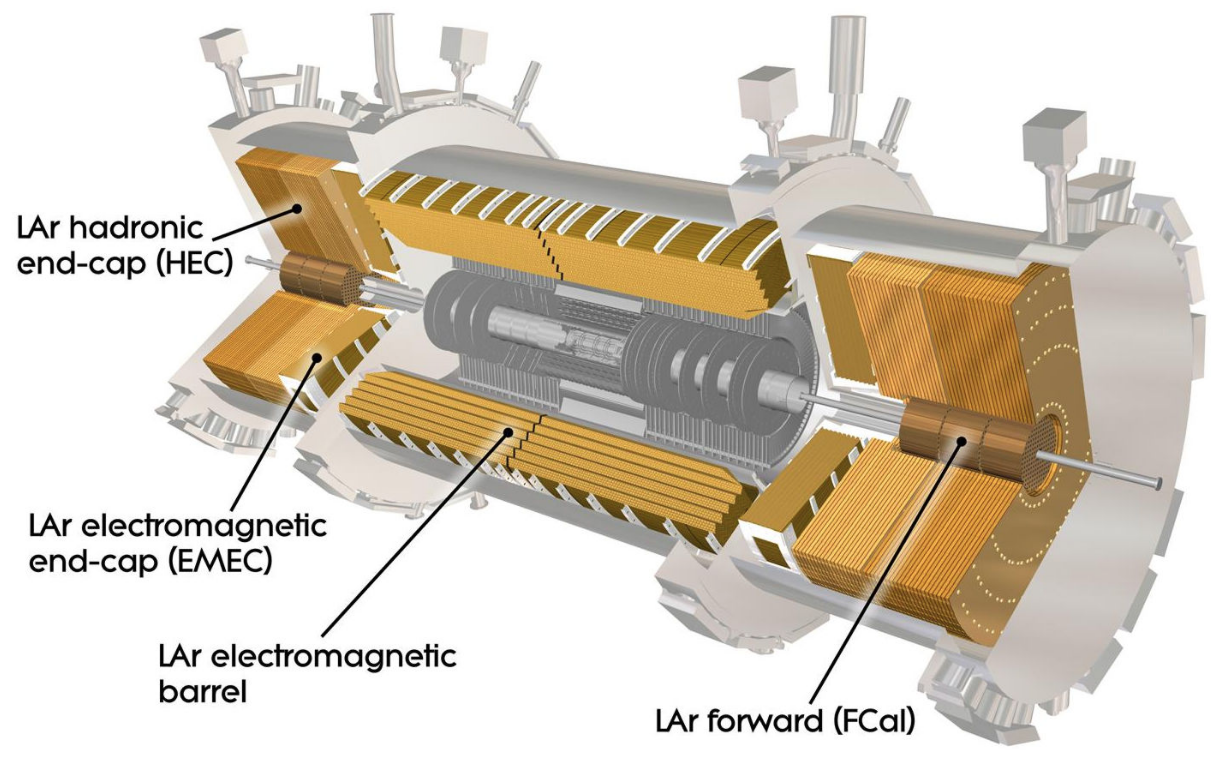}
	\caption{ATLAS Liquid Argon calorimeter}
	\label{lar}
\end{figure}

\begin{figure}[h!]
	\centering
	\includegraphics[width=0.95\linewidth]{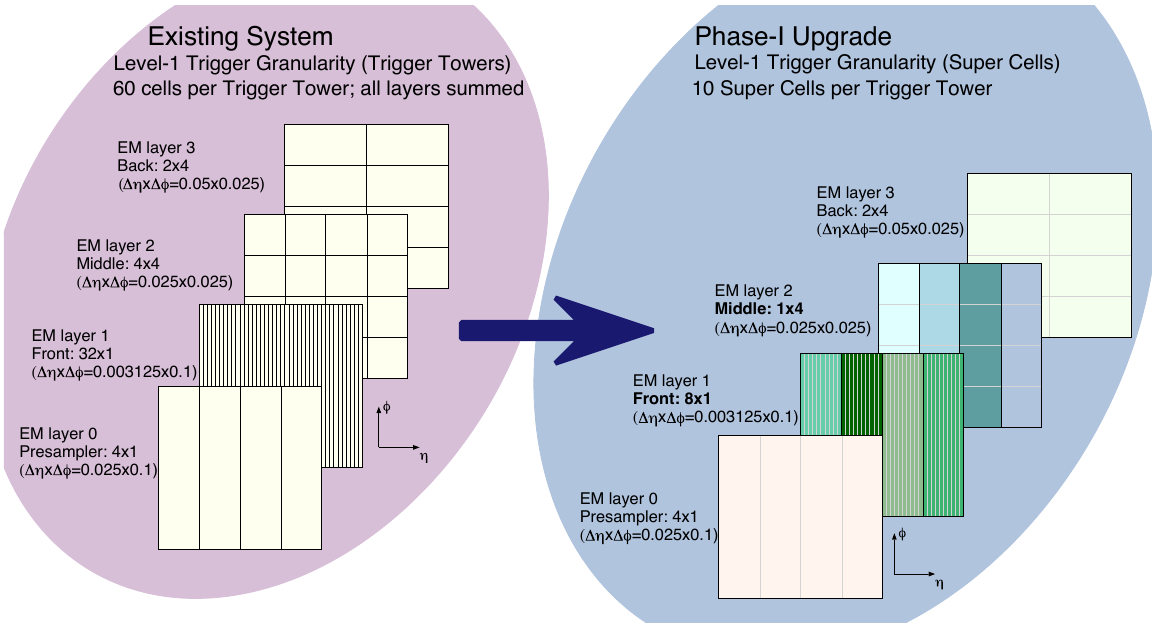}
	\caption{Geometrical representation in $\eta$,$\phi$ space of an electromagnetic trigger tower in the current system, where the transverse energy in all four layers are summed (left) and of the super-cells for the Phase-I upgrade, where the transverse energy in each layer is retained in addition to the finer granularity in the front and middle layers (right). Each big square here represents an area of size $\Delta\eta\times\Delta\phi = 0.1\times0.1$.\cite{atlas-tdr-2}}
	\label{geometrical}
\end{figure}

ATLAS Liquid Argon (LAr) calorimeter consists of the electromagnetic barrel, the electromagnetic end-cap, the hadronic end-cap and the forward calorimeter\cite{atlas-1}\cite{atlas-tdr-1}. The position of these calorimeters are shown in the Figure\,\ref{lar}. In current LAr trigger readout, for each area of size $\Delta\eta\times\Delta\phi = 0.1\times0.1$, the Layer Sum Board (LSB), as mezzanines on the front-end board will sum and get the energy deposition across each of the four longitudinal layers in the calorimeter. As depicted in Figure\,\ref{geometrical}, the so called Tower Builder Board will further sum these four energies together, and form a trigger tower with the granularity of $0.1\times0.1$.

The second long shutdown of LHC is scheduled in 2019-2020. For LAr calorimeters of the ATLAS detector, the Phase-I upgrade of the trigger readout electronics will be installed. The objective of this upgrade is to provide higher granularity, higher resolution and longitudinal shower information\cite{atlas-tdr-2}. As shown in the Figure\,\ref{geometrical}, after the upgrade, the level-1 trigger granularity will be improved. One current trigger tower will has 10 so called super cells. The information from each layer is retained, and the granularity can be fine up to $\Delta\eta\times\Delta\phi = 0.025\times0.1$. There will be about 34000 super cells in total. All of them will be sampled at every LHC bunch-crossing at a frequency of 40 MHz.

As the LHC luminosity increases above the LHC design value, the improved calorimeter trigger electronics will allow ATLAS to deploy more sophisticated algorithms, enhancing the ability
to measure the properties of the newly discovered Higgs boson and the potential for discovering physics beyond the standard model.

\begin{figure}[h!]
	\centering
	\includegraphics[width=1\linewidth]{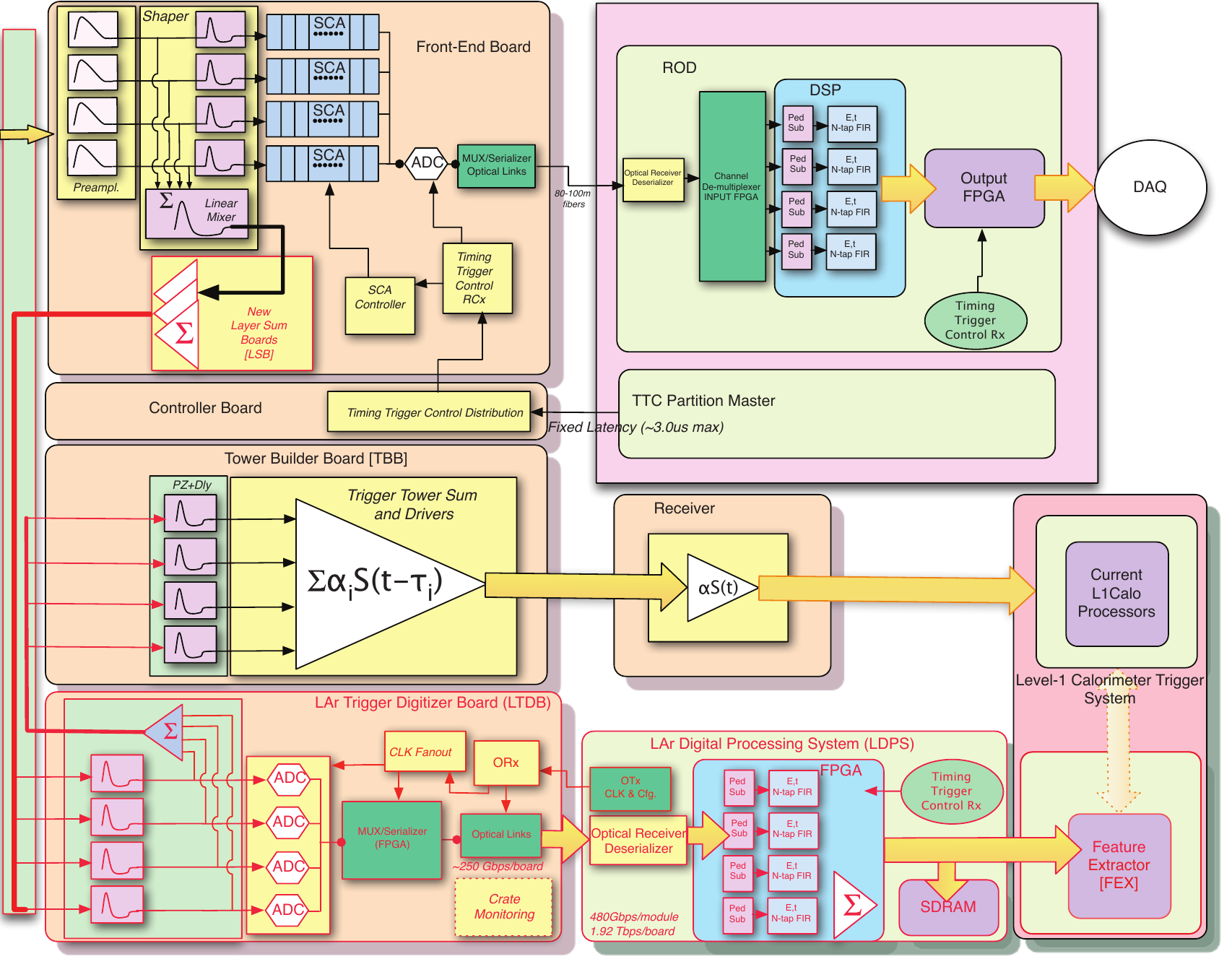}
	\caption{Schematic block diagram of the upgraded LAr trigger readout architecture: with new components indicated by the red outlines and arrows.}
	\label{fig:larph1dg}
\end{figure}

As shown in the architecture of the Figure\,\ref{fig:larph1dg}, LSBs need to be upgraded to output the super cell signals. The new LAr Trigger Digitizer Boards (LTDB) will process and digitize the super cell signals, and send the processed data to the back-end electronics LAr Digital Processing System (LDPS), where data are then transmitted to the trigger processors. Each LTDB will be able to process up to 320 super cell signals, which will be digitized by 80 of 12-bit quad-channel NEVIS ADCs\cite{adc} on the LTDB. Twenty serializer ASICs LOCx2 will receive the ADC data, and output 40 of 5.12 Gb/s data via fiber optical links to the LDPS\cite{locx2}. The LDPS will convert the samples to calibrated energies in real-time and interface to the FEX processors. With a total of 124 LTDB boards in the system, the total rate to the back end electronics is approximately 25 Tbps. The LTDB will also output 64 summed signals to the tower builder board, each of the signal is sum of 4 input in the same middle or front layer.

The control and monitoring of LTDB is realized via 5 GBT links connected with the Front-End LInks eXchange (FELIX) in ATLAS TDAQ system\cite{felix-1}\cite{felix-2}. FELIX will distribute the TTC (Time, Trigger and Control) clock and BCR (Bunch Counter Reset) signal via down-links to the LTDB. Besides TTC information, FELIX will also control GBTx and all ASICs via the SCA (Slow Control Adapter) ASIC on the LTDB.

\section{Design and Test}
\subsection{Design of the LTDB}
\begin{figure}[H]
	\centering
	\includegraphics[width=0.9\linewidth]{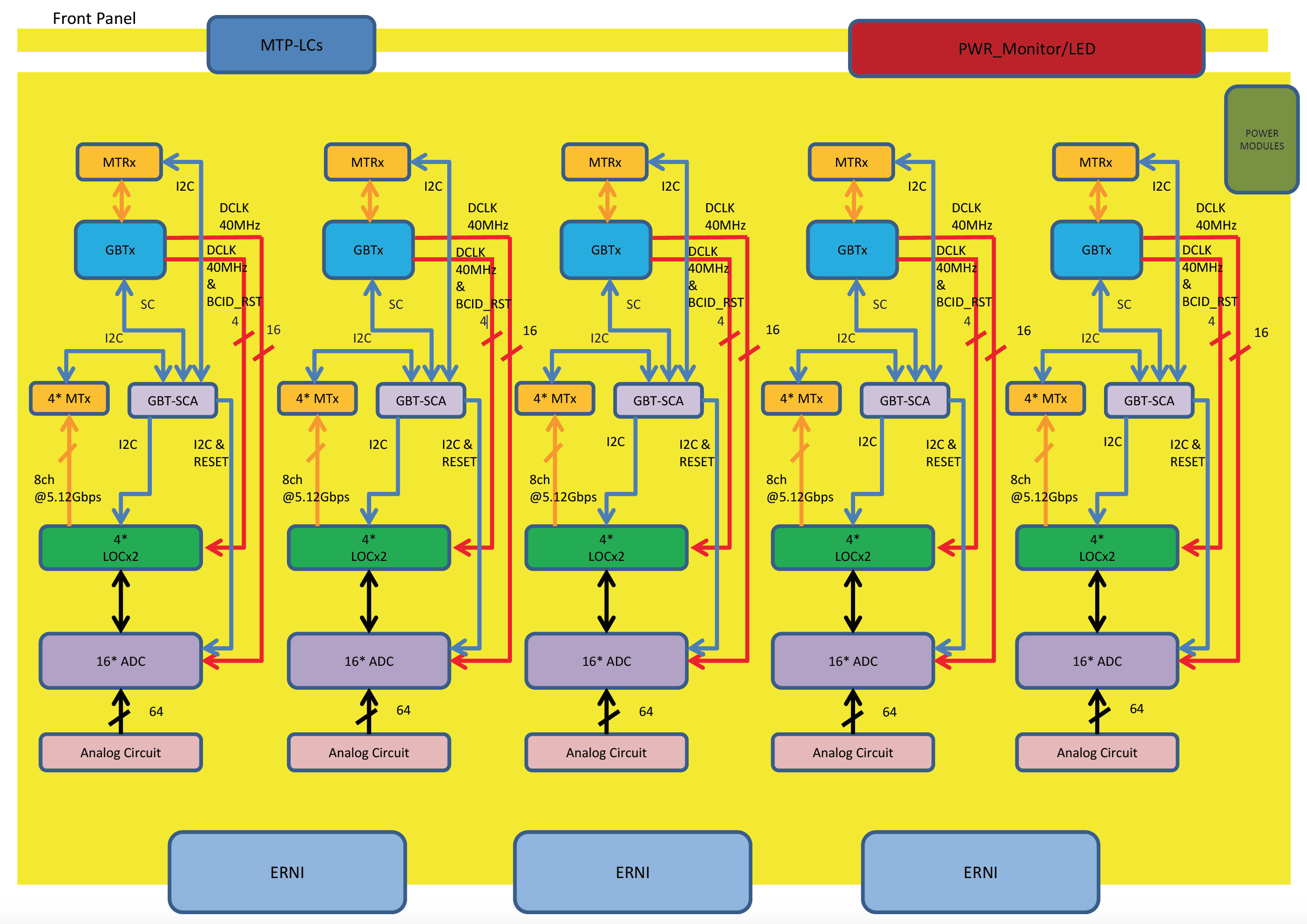}
	\caption{Diagram of the LTDB prototype}
	\label{fig:diagram}
\end{figure}

\begin{figure}[H]
	\centering
	\includegraphics[width=0.7\linewidth, angle=90]{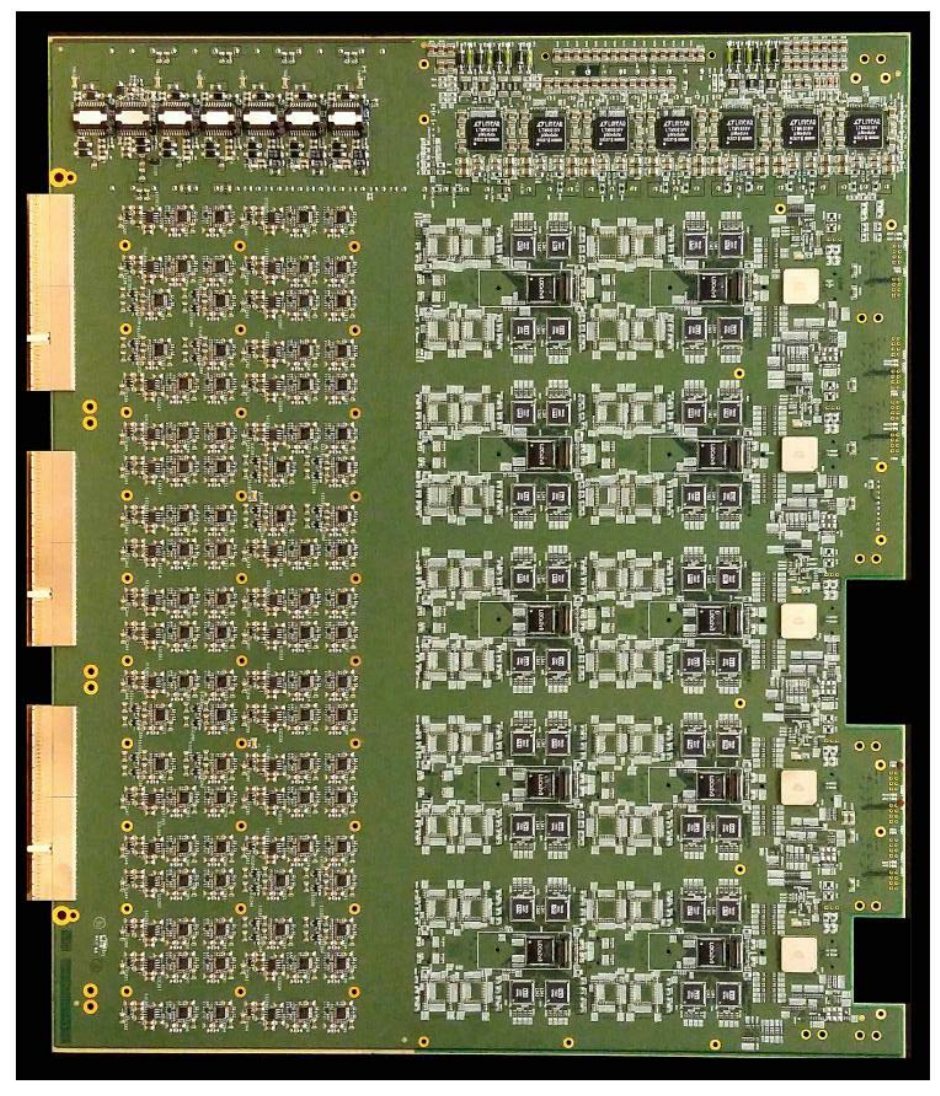}
	\caption{LTDB prototype}
	\label{fig:board}
\end{figure}

The LTDB design have been split into three stages. In the demonstrator stage, forty 8-channel TI ADC ADS5272 digitize the 320 super cells. Four Kintex-7 FPGA are used to packing the data and send it to the back-end via 40 of 4.8 Gb/s links. Radiation tolerance of COTS ADCs and power converters are researched\cite{adctest-1}\cite{adctest-2}. For the pre-prototype, 80 of the NEVIS ADCs are used, 10 of Xilinx Artix-7 FPGA are used for data packing and transmission. From the prototype stage, all of the ADC, serilizer, optical-electric converters are custom radiation-hard ASICs. Diagram of the LTDB prototype is shown in the Figure\,\ref{fig:diagram}.
The prototype board is shown in Figure\,\ref{fig:board}. Analog section is at the bottom half of board. Digital part is at the top side.

The slow control between FELIX and LTDB are realized via 5 GBT links. The GBT link is connected to GBTx via the MTRx module. GBTx supports to output two kinds of recovered clock: DCLK with better quality and CLKDES with programmable phase in step of 48.8 ps. For 40MHz, the jitter in a frequency range from 100 Hz to 5MHz is about 4 ps for DCLK. For CLKDES, it is about 10 ps. The high quality DCLK is used as the ADC input clock. On the prototype board, LOCx2 also uses DCLK. On the pre-production board, the CLKDES\cite{bibGBTx} is used to support the phase calibration required by the LOCx2. 

\subsection{Test Setup and Results}

\begin{figure}[h]
	\centering
	\includegraphics[width=0.9\linewidth]{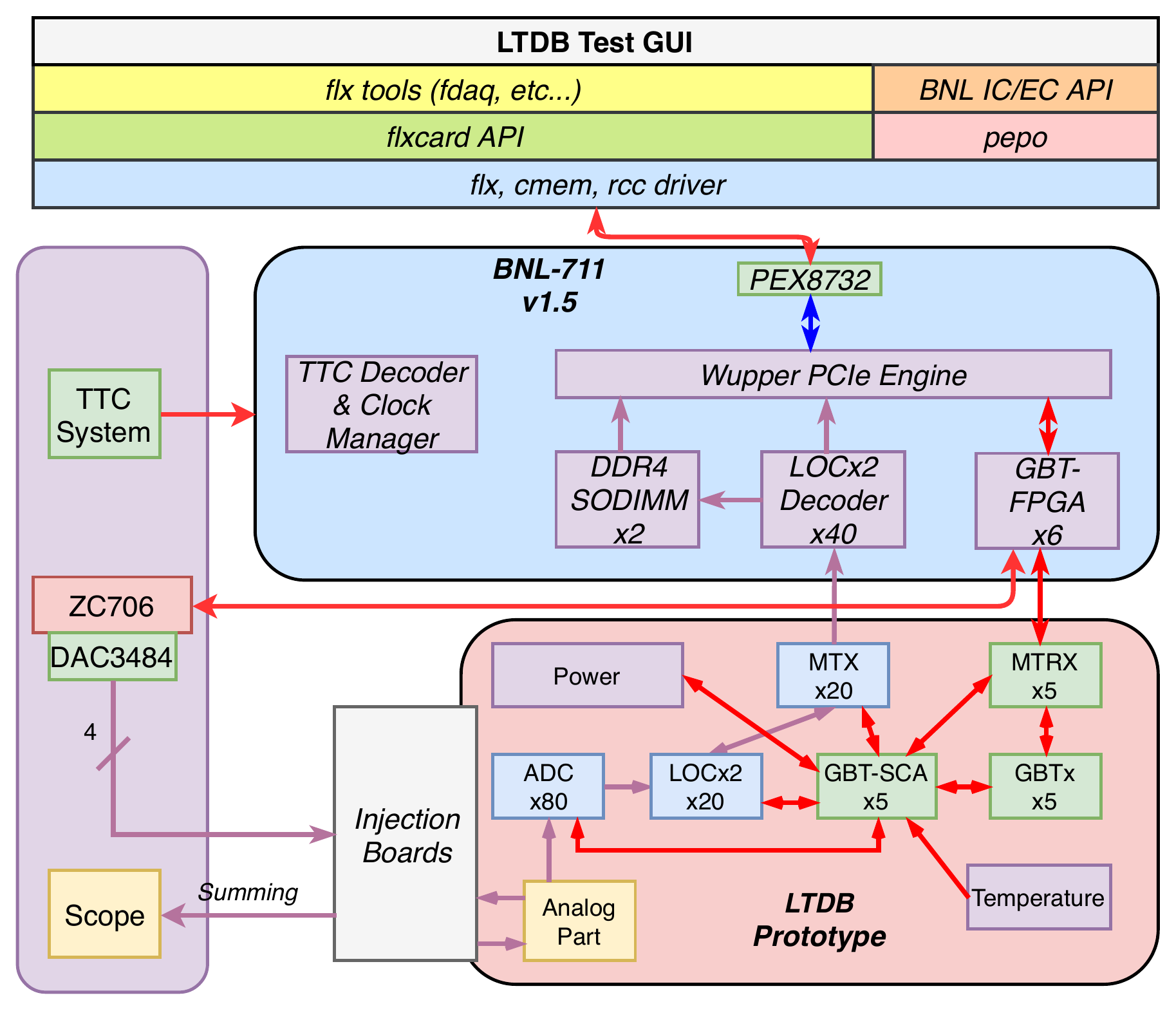}
	\caption{Test setup for LTDB evaluation}
	\label{fig:testsetup}
\end{figure}

To test the LTDB boards, as shown in the test setup of Figure\,\ref{fig:testsetup}, the PCIe card BNL-711 is developed. This 16-lane Gen 3 PCIe card can interface the ATLAS TTC system, and decode the TTC clock and TTC information like BCR. There are 48 bi-directional optical links which can run up to 14 Gb/s. The two DDR4 modules can support 32 GB buffer with a speed of up to about 270 Gbps. 

\begin{figure}[H]
	\centering
	\includegraphics[width=0.9\linewidth]{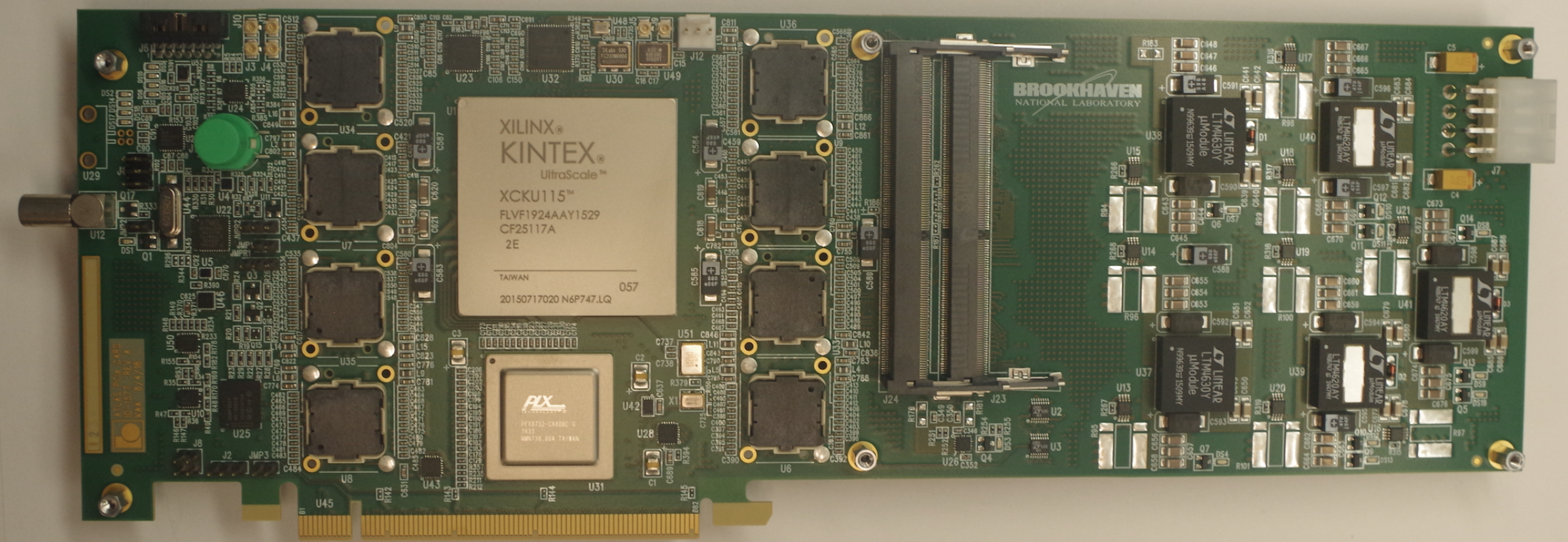}
	\caption{BNL-711 PCIe card for the LTDB test}
	\label{fig:BNL-711}
\end{figure}

\begin{figure}[H]
	\centering
	\includegraphics[width=0.9\linewidth]{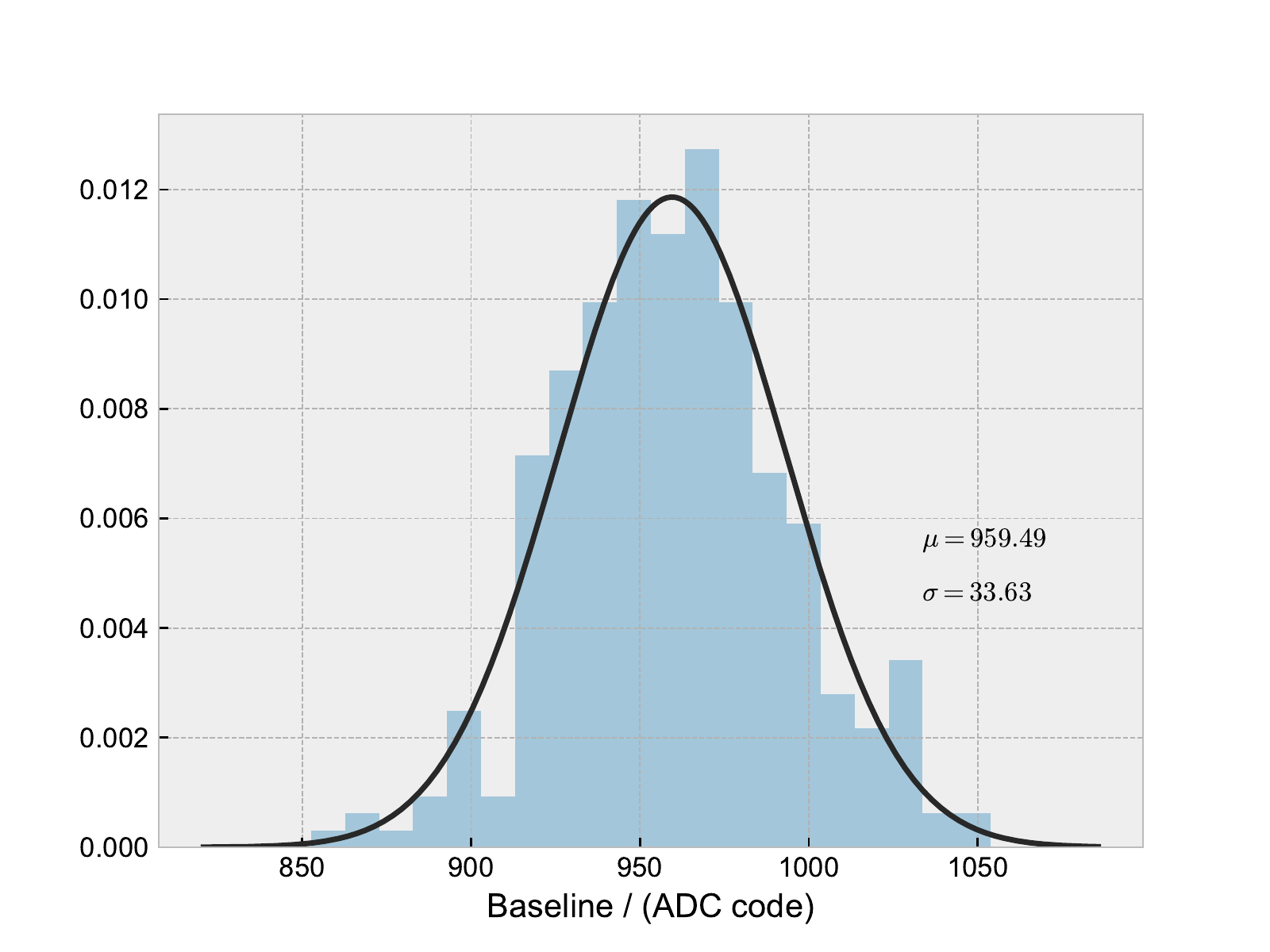}
	\caption{Pedestal distribution of the 320 channels}
	\label{fig:pedestal}
\end{figure}

\begin{figure}[H]
	\centering
	\includegraphics[width=0.9\linewidth]{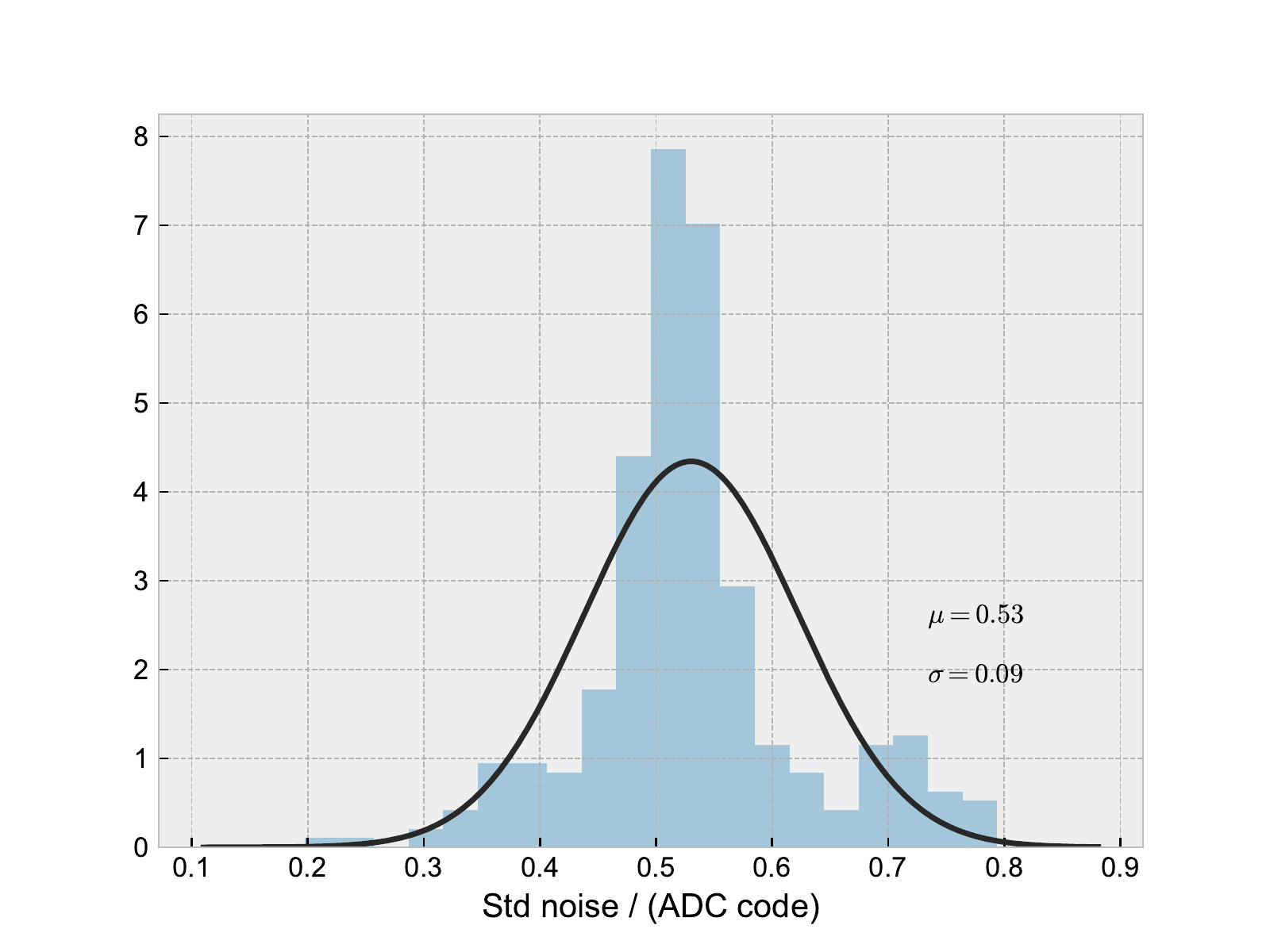}
	\caption{Std noise distribution of the 320 channels}
	\label{fig:noise}
\end{figure}

In the test setup, the firmware has 6 bidirectional GBT links and 40 GTH receiver for the data links from the LTDB. The low-level software tools from the FELIX project are used to read and write registers in the Kintex Ultrascale FPGA on BNL-711. An optimized low latency GBT-FPGA core is developed to support multi-channel GBT links with LTDB\cite{opt-gbt}. The IC/EC API is developed to control the GBTx and communicate with GBT-SCA chips via the GBT links\cite{HDLC-ICEC}. With this API, all ASICs on the LTDB can be configured and calibrated. The high level software is designed to monitor voltage, current and temperature of the LTDB. Beside the five GBT links for LTDB, one extra link is connected to the Xilinx ZC706 evaluation board. This link is used control the DAC3484 evaluation card, which can generate test super cell waveform for LTDB. The Xilinx all digital VCXO PICXO\cite{VCXO} is implemented in ZC706 to synchronize the GBT link to BNL-711 with system clock from BNL-711. The BNL-711 PCIe card are shown in Figure\,\ref{fig:BNL-711}.
With this test setup, all functions of the LTDB pre-production are verified. With the successful testing, two LTDB pre-production boards are installed on the detector in early 2018. The pedestal and noise distribution of all channels on one board are shown in the Figure\,\ref{fig:pedestal} and \ref{fig:noise}. 

\section{Summary and outlook}
The new LAr calorimeter trigger readout system is being designed for the Phase-I upgrade. LTDB is the kernel part in the front-end. Two LTDB pre-production boards have been installed on the detector in 2018 successfully.
The preliminary test results show that the total noise level of the crate with LTDB installed is at the same level of other crates. The pedestal and noise level of super cells are same with test at lab. The data taking is ongoing for the LHC Run 2 with the BNL-711 PCIe card. Full integration with LDPS, FELIX and level 1 calorimeter trigger system is scheduled in summer of 2018.






%

\end{document}